\documentstyle[12pt]{article}
\newcommand{\uq}{U_q(gl(M|N))}
\newcommand{\gl}{gl(M|N)}
\newcommand{\Yq}{Y_q(gl(M|N))}
\newcommand{\gt}{+}
\newcommand{\ba}{\begin{eqnarray}}
\newcommand{\na}{\end{eqnarray}}
\newcommand{\ban}{\begin{eqnarray*}}
\newcommand{\nan}{\end{eqnarray*}}
\newtheorem{lemma}{Lemma}

\newcommand{\x}{\otimes}

\begin{document}
\title{\bf{\small THE QUANTUM SUPER YANGIAN AND \\
          \small  CASIMIR OPERATORS OF $\uq$ \\} }
\author{\small R. B. ZHANG\\
\small Department of Pure Mathematics\\
\small University of Adelaide \\
\small Adelaide, Australia}
\date{}
\maketitle

\vspace{3cm}
\small
The ${\bf Z}_2$ graded Yangian $\Yq$ associated with the Perk - Schultz
$R$  matrix is introduced. Its structural properties, the central algebra
in particular,  are studied.  A ${\bf Z}_2$ graded associative algebra
epimorphism $\Yq\rightarrow\uq$ is obtained in explicit form.  Images
of central elements of the quantum super Yangian under this epimorphism
yield the Casimir operators of the quantum supergroup $\uq$ constructed
in an earlier publication.

\vspace{2cm}
\normalsize
\noindent
\section{\small INTRODUCTION}
Quantum supergroups\cite{sgroups} are a particularly interesting
class of deformations of the universal enveloping algebras of
the basic classical Lie superalgebras. Their origin may be traced
back to the Perk -- Schultz solution\cite{perk} of the Yang -- Baxter
equation, and the work of Bazhanov and shadrikov\cite{vvb} on
the $R$ -- matrices associated to the vector representation
of $osp(M|2N)$(For a quantum supergroup interpretation of these $R$ --
matrices, see \cite{PLB}). However, systematic studies of these algebraic
structures only began very recently, largely motivated by
applications to soluble lattice models in statistical mechanics and
knot theory. Now the understanding of the so -- called type I quantum
supergroups is rather complete, but many important problems remain open
for the type II quantum supergroups.

Another kind of algebraic structure closely related to the quantum
supergroups is quantum super Yangians, which are generalizations of
Drinfeld's Yangian algebras\cite{drinfeld}, and also arise naturally
from the algebraic description of soluble lattice models in
statistical mechanics.  The classical super Yangians have been studied
by Nazarov\cite{sy}, who in particular explored the hidden super Yangian
symmetry of the classical Lie superalgebras. In this letter we will
study the structural features of the quantum super Yangian associated
with the Perk -- Schultz $R$ matrix, and also to investigate
the interrelationship between this quantum super Yangian and
the quantum supergroup $\uq$, thus to better our understanding of the
structures of the latter from a Yangian point of view.
We wish to point out  that Yangians provide a new framework for studying the
ordinary Lie algebras, within which many of the structural and
representation theoretical problems can be treated in a unified manner.
For example, the recent Yangian interpretations of the Gelfand - Zetlin
bases for irreducible modules\cite{gz} and the central algebra of $gl(M)$
are fascinating developments in the theory of Lie algebras(for a review
see \cite{molev}).

We will define the quantum super Yangian $Y_q(\gl)$ associated with
the Lie superalgebra $\gl$  employing the Faddeev --
Reshetikhin -- Takhtajian approach to quantization of algebras,
starting from the Perk -- Schultz solution of the Yang -- Baxter
equation. The Hopf algebraic structure of the quantum super Yangian
will be elucidated, and a set of central elements will be constructed
explicitly.
We will also demonstrate that there exists a homomorphism
of ${\bf Z}_2$ graded associative algebras from $Y_q(\gl)$ to $\uq$,
which is surjective, and explicitly realized by the universal $L$ --
operator constructed in \cite{I}. Images of the $Y_q(\gl)$ central
elements under this epimorphism are Casimir operators of $\uq$,
which turn out to be those constructed in \cite{I} using the $q$ --
(super)trace technique developed in  earlier
publications\cite{II}\cite{III}.

As the ordinary quantum group $U_q(gl(M))$ is nothing esle but
$\uq$ at $N=0$, $\Yq$ at $N=0$ reduces to the quantum Yangian studied
in \cite{qy}. Our results on central elements also  yield, in this case,
a Yangian interpretation of the $U_q(gl(M))$ Casimir operators
of \cite{II}, which will be shown to generate the entire center of
$U_q(gl(M))$.

\section{\small VECTOR REPRESENTATIONS OF $\uq$}
We will largely follow the notation of \cite{I}, but adopt a different
convention for the co - multiplication etc.. To avoid confusion, we present
the definition of the quantum supergroup $\uq$ here. We will also briefly
explain the $R$ - matrices associated with its vector and dual vector modules,
which are of crucial importance for studying the quantum super Yangian
$\Yq$.

The quantum supergroup $\uq$  is a ${\bf Z}_2$-graded unital algebra
generated by $E^a_{a\pm 1}$, $E^a_a$, $a$, $a\pm 1 = 1, 2,...,M+N$,
subject to the following relations
\ba
[E^a_a,\,E^b_b\} =0,  &   {[}E^a_a,\,E^b_{b\pm 1}\} = (\delta^b_a
- \delta^a_{b\pm 1}) E^b_{b\pm 1}, \ \ \ \forall a, b\nonumber\\
{[}E^a_{a+1},\,E^{b+1}_b\} = (-1)^{[a]}\delta _{ab}
{{q^{h_a} - q^{-h_a}}\over {q - q^{-1}} },  &
(E^m_{m+1})^2 = (E^{m+1}_m)^2 = 0,  \nonumber\\
E^a_{a+1}\,E^b_{b+1} = E^b_{b+1}\,E^a_{a+1},&
E^{a+1}_a\,E^{b+1}_b = E^{b+1}_b\,E^{a+1}_a, \ \ \  |a - b|\geq 2, \nonumber
\na
\ba
(E^a_{a+1})^2\,E^{a\pm 1}_{a\pm1 +1} - (q +
q^{-1})\,E^a_{a+1}\,E^{a\pm 1}_{a\pm 1+1}\,E^a_{a+1} +
E^{a\pm 1}_{a\pm 1+1}\,(E^a_{a+1})^2 = 0,&\  \ \  a \neq m,\nonumber\\
(E^{a+1}_a)^2\,E^{a\pm 1 + 1}_{a\pm 1} - (q +
q^{-1})\,E^{a+1}_a\,E^{a \pm 1 + 1}_{a \pm 1}\,E^{a +
1}_a + E^{a\pm 1 + 1}_{a\pm 1}\,(E^{a+1}_a)^2 =0, &\ \ \   a\neq m,
\nonumber\\
{[}E^{m-1}_{m+2},\,E^m_{m+1}\} = [E^{m+2}_{m-1},\,E^{m+1}_m\} = 0,&
\na
where
 $$h_a =(-1)^{[a]}\,E^a_a - (-1)^{[a+1]}\,E^{a+1}_{a+1},$$
with $[a] =0$ for $a\leq m$ and $[a] =1$ for $a > m$;
$[.,\,.\}$ represents the standard graded commutator;
and $E^{M-1}_{M+2}$, $E^{M+2}_{M-1}$ are the $a=M-1$,
$b=M+1$ case of the following elements
\ba
E^a_b &=& E^a_c\,E^c_b - q^{-(-1)^{[c]}}\,E^c_b\,E^a_c,\nonumber \\
E^b_a &=& E^b_c\,E^c_a - q^{(-1)^{[c]}}\,E^c_a\,E^b_c, \ \ \ \
  a < c < b. \label{E}
\na
Throughout the paper, we assume that $q\in \bf C$ is not a root of unity.

The co-multiplication $\Delta :
\uq \rightarrow \uq\x \uq$ is taken to be
\ban
\Delta(E^a_{a+1}) &=& E^a_{a+1} \otimes 1 + q^{h_a}\otimes E^a_{a+1},\\
\Delta(E^{a+1}_a) &=& E^{a+1}_a \otimes q^{-h_a} + 1 \otimes E^{a+1}_a,\\
\Delta(E^a_a) &= &E^a_a \otimes 1 + 1 \otimes E^a_a;
\nan
and the co - unit $\epsilon: \uq\rightarrow {\bf C}$ is defined by
\ban
\epsilon(E^a_{a\pm 1}) &=  \epsilon (E^a_a) &= 0,\\
\epsilon (1) &= 1.
\nan
Then the corresponding  antipode $S :\uq \rightarrow \uq$ is given by
\ban
S(E^{a}_{a+1}) &=& - q^{-h_a}E^a_{a+1},\\
S(E^{a+1}_a) &=& - E^{a+1}_a q^{h_a},\\
S(E^a_a) &=& - E^a_a,
\nan
which is a ${\bf Z}_2$-graded algebra anti-automorphism, i.e.,
for homogeneous elements $x$, $y \in \uq$, we have
$S(x y) = (-1)^{[x][y]}\,S(y)\,S(x),$
and generalizing to inhomogeneous elements through linearity.
We will denote the opposite co-multiplication by $\Delta'$.

The vector module over $\uq$ is a ${\bf Z}_2$-graded vector space $V$
of dimension $M+N$, for which we choose a basis
$\{ v^a \ | \ a=1,2,..., M+N\}$,
that is homogeneous with $v^a$ being evn if $[a]=0$ and odd if $[a]=1$.
The action of $\uq$ is now defined by $E^{a}_{b} v^c = \delta_b^c v^a$,
for $b= a-1,\ a, \ a+1$. We denote the associated vector representation of
$\uq$ by $\pi$. Then in this basis $\pi(E^{a}_{b})=e^{a}_{b}$, where
$e^{a}_{b}\in End(V)$ are the standard matrix units.

We denote the dual vector module over $\uq$ by $V^*$, and the corresponding
representation by $\pi^*$. A useful basis for $V^*$ is
$\{ v^*_a \ | \ a=1,2,..., M+N\}$ defined by $v^*_a(v^b)=\delta_a^b$.
The action of  $\uq$ is defined by $E^{a}_{b} v^*_c (v^d)=
(-1)^{([a]+[b])[c]}\times $   $v^*_c(S(E^a_b) v^d)$.
Let $^\gt$ denote the graded transposition on matrices
defined by $(e^a_b)^\gt=$  $(-1)^{([a]+[b])[a]}e^b_a$.
Then we have $\pi^*(E^a_{a\pm 1})
=-q^{\mp 1} (e^a_{a\pm 1})^\gt$, $\pi^*(E^a_a)=-e^a_{a}$.

The Perk - Schultz solution of the Yang - Baxter equation is associated
with the tensor product module $V\x V$ over $\uq$, and is given by
\ban
R(x)&=& q^{\sum_a e^a_a\otimes e^a_a(-1)^{[a]}}
     - x^{-1}q^{-\sum_a e^a_a \otimes e^a_a (-1)^{[a]}}\\
    &+& (q - q^{-1})(x^{-1}\sum_{a<b}+\sum_{a>b})e^a_b
     \otimes e^b_a (-1)^{[b]}, \ \ \ x\in {\bf C}.
\nan
For later use we also give the solution $R(x)^*$ of the Yang - Baxter
equation associated with the $\uq$ module $V^*\x V$, which can be
expressed in terms of $R(x)$ through
\ban
R^*(x)&=&(R(x)^{-1})^{\gt_1}(q-x^{-1}q^{-1})(q^{-1}-x^{-1}q),
\nan
where $^{\gt_1}$ denotes the graded transposition over the first space,
and the numerical factor is introduced for convenience.  Explicitly,
\ban
R^*(x)&=& x^{-1}\left[q^{\sum_a e^a_a\otimes e^a_a(-1)^{[a]}}
     +(q - q^{-1})\sum_{a<b} e^b_a\x e^b_a (-1)^{[a][b]+[a]+[b]}\right]\\
     &-& \left[q^{-\sum_a e^a_a \otimes e^a_a (-1)^{[a]}}
   - (q - q^{-1})\sum_{a>b}e^a_b \otimes e^b_a (-1)^{[a][b]+[a]+[b]}\right] .
\nan
$R^*$ satisfies the following equations
\ba
R^*_{12}(x) R^*_{13}(x y) R_{23}(y)&=&
R_{23}(y) R^*_{13}(x y) R^*_{12}(x),\nonumber\\
R^*(x)(\pi^*\x\pi)\Delta(u)&=&(\pi^*\x\pi)\Delta'(u)R^*(x),
\ \ \ \forall u\in\uq.
\na

It is important to observe that $V^*\x V$ can be decomposed into
the direct sum of two irreducible $\uq$ modules,$V^*\x V=A\oplus
\{|0\rangle\}$, with $\{|0\rangle\}$ being trivial, and $A$
a quantum deformation of the adjoint module of $\gl$.
The tensor product $V\x V^*$ also has these properties, namely,
$V\x V^*={\bar A}\oplus \{|{\bar 0}\rangle\}$. Here the modules are
defined with respect to the co - multiplication $\Delta$.
We introduce the projection operators
\ban
P[A](V^*\x V)={\bar A},& {\bar P}[A](V\x V^*)=A,\\
P[0](V^*\x V)=\{|{\bar 0}\rangle\}, & {\bar P}[0](V\x V^*)= \{|0\rangle\},\\
P[A]+P[0]=P,& {\bar P}[A] + {\bar P}[0] = {\bar P},
\nan
where $P:V^*\x V\rightarrow V\x V^*$ and
${\bar P}: V\x V^* \rightarrow  V^*\x V$ are the graded permutation
operators respectively defined by
$P(v^*_a\x v^b)=(-1)^{[a][b]}v^b\x v^*_a$,
and ${\bar P}(v^b\x v^*_a)=$ $(-1)^{[a][b]}v^*_a\x v^b$.
Needless to say, these projection operators all commute with $\Delta(u)$,
$\forall u\uq$, e.g., $(\pi\x\pi^*)\Delta(u)P[A] = P[A](\pi^*\x\pi)\Delta(u)$,
and also satisfy the relations $P[0]{\bar P}[A]=P[A]{\bar P}[0]=0$,
$P[0]{\bar P}[0]+P[A]{\bar P}[A]=id_{V\x V^*}$, and similar relations obtained
by exchanging the orders of the two sets of operators.
These relations define the projectors uniquely.

In terms of the projectors and the permutation operators, $R$ and $R^*$
can be expressed as
\ba
PR^*(x)&=&(1-x^{-1})P[A] + (q^{N-M}-x^{-1}q^{M-N})P[0], \nonumber\\
(R^*(x))^{-1}{\bar P}&=&{ {{\bar P}[A]}\over {1-x^{-1}}}
    +{ { {\bar P}[0]}\over {q^{N-M}-x^{-1}q^{M-N}} }.\label{R}
\na
For the purpose of this letter, we will only need explicit matrix forms of
the projection operators $P[0]$ and ${\bar P}[0]$, which we spell out below
\ban
P[0]&=&{ {1}\over {SD_q(V)}} \sum_{a, b} (-1)^{[b]([a]+[b])} e^a_b\x e^a_b,\\
{\bar P}[0]&=& { {1}\over {SD_q(V)}} \sum_{a, b} (-1)^{[a]([a]+[b])}
\pi(q^{h_{2\rho}}) e^a_b\pi(q^{-h_{2\rho}})\x e^a_b,
\nan
where $h_{2\rho}$ is the linear sum of $h_a$'s such that
\ban
[h_{2\rho}, E^a_{a+ 1}] =2(-1)^{[a]}(1-\delta_{a m}) E^a_{a+ 1}, &
[h_{2\rho}, E^{a+1}_a]=-2(-1)^{[a]}(1-\delta_{a m}) E^{a+1}_a,
\nan
and $SD_q(V)$ is the $q$ - superdimension of $V$.

\section{\small THE QUANTUM SUPER YANGIAN AND ITS CENTRAL ALGEBRA}
With the above preparations, we can now define $\Yq$, the quantum super
Yangian associated with $\gl$,  and investigate its properties.
$\Yq$ is a ${\bf Z}_2$ graded algebra generated by
$\{ T^a_b[k] \ | \ 0<k\in{\bf Z}, \ a, b = 1,2,..., M+N\}$, with the following
quadratic relations
\ba
R_{12}(x) L_1(x y) L_2(y)&=& L_2(y) L_1(x y) R_{12}(x),\label{definition}
\na
where
\ban
L(x)=\sum_{a, b}(-1)^{[b]}e^a_b\x T^b_a(x),&
T^b_a(x)=(-1)^{[b]}\delta^b_a + \sum_{0<k\in{\bf Z}_+} x^{-k} T^b_a[k].
\nan
The element $T^a_b[k]$ is assumed to be even if $[a]+[b]\equiv 0(mod 2)$,
and odd otherwise.

This algebra admits co - algebra structures compatible with
the associative multiplication defined by equation (\ref{definition}).
Explicitly, there exist a co - unit $\sigma: \Yq\rightarrow {\bf C}$,
$T^a_b[k]\mapsto\delta_{0, k}\delta^a_b(-1)^{[a]}$, and a co -
multiplication ${\hat\Delta} :   \Yq\rightarrow$
$ \Yq\x$ $\Yq$, $L(x)\mapsto L(x)\x L(x)$.
An antipode $\gamma: \Yq\rightarrow \Yq$,
$L(x)\mapsto L^{-1}(x)$ can also be introduced, thus turning $\Yq$ into a
${\bf Z}_2$ graded Hopf algebra. Note that matrix elements of
$\gamma(L(x))$ $=1\x 1 +\sum_{0<n\in{\bf Z}_+} (1\x 1-L(x))^n$
are in $\Yq$, and well defined on ${\bf C}[[x^{-1}]]$
under a $p$ -adic type topology.

For convenience, we write
\ban
\gamma(L(x))=\sum_{a, b}(-1)^{[b]}e^a_b\x {\tilde T}^b_a(x), &
{\tilde T}^b_a(x)=\sum_{n\in{\bf Z}_+}x^{-1} {\tilde T}^b_a[n],
\ \ \ {\tilde T}^b_a[0]=(-1)^{[b]}\delta^b_a,
\nan
and define
\ban
L^*(x)&:=(\gamma(L(x)))^\gt&
       =\sum_{a, b}(-1)^{[a][b]+[a]+[b]}e^b_a\x{\tilde T}^b_a(x).
\nan
Then in terms of $L^*(x)$, the defining relations for $\Yq$
can be rewritten as
\ba
R^*_{12}(x) L^*_1(x y) L_2(y)&=& L_2(y) L^*_1(x y) R^*_{12}(x). \label{redef}
\na
Pre and post multiplying  (\ref{redef}) by $(R^*_{12}(x))^{-1}$ results
in  an equation, which has simple poles at $x=1$ and $x=q^{2(M-N)}$, as can be
easily seen by recalling the explicit form of $(R^*(x))^{-1}$ given in
(\ref{R}). Extracting the residue of this equation at  $x=q^{2(M-N)}$,
we arrive at
\ba
{\bar P}_{12}[0] P_{12} L_2(y) L_1^*(y q^{2(M-N)}) {\bar P}_{12}
&=& L_1^*(y q^{2(M-N)})  L_2(y) {\bar P}_{12}[0], \label{main}
\na
which, in explicit form,  reads
\ban
& &\sum_{a, r, s} \pi(q^{h_{2\rho}})e^a_r\x e^a_s\x
\sum_b \left(\pi(q^{-h_{2\rho}})\right)^b_b T^r_b(y){\tilde T}^b_s(y
q^{2(M-N)})
(-1)^{[b]+\zeta}\\
&=& \sum_{b, r, s} e^r_b\pi(q^{-h_{2\rho}})\x e^s_b
\x \sum_a  \left(\pi(q^{h_{2\rho}})\right)^a_a{\tilde T}^r_a(y
q^{2(M-N)})T^a_s(y)
(-1)^{[a]+\eta},
\nan
with
\ban
\zeta&\equiv& [r]+[s]+[r][s]+[a]([a]+[r])(mod \ 2),\\
\eta&\equiv& [r] + [s]([r]+[b])(mod \ 2).
\nan

Denote the the right hand side of (\ref{main}) by $RHS_{(\ref{main})}$.
We have
\ban
P_{12}[0]RHS_{(\ref{main})}P_{12}[0] &=& P_{12}[0]\x
str_\pi[\pi(q^{h_{2\rho}}) L^{-1}(y q^{2(M-N)})L(y)].
\nan
Define
\ba
C(x)&=&str_\pi[\pi(q^{h_{2\rho}}) L^{-1}(x q^{2(M-N)})L(x)] \ \in\Yq.
\na
We claim that
\begin{lemma}
$C(x)$ can be expanded into
\ban
C(x)&=&\sum_{n\in{\bf Z}_+} x^{-1}C_n,
\nan
with  $C_n\in\Yq$, $n=0, 1,...$, belonging to the center of $\Yq$, namely,
\ba
[C_n, T^a_b[k]]&=&0, \ \ \ \forall a, b, k, n.
\na
\end{lemma}
{\em Proof}: Let us examine the following equation
\ba
& &L^*_1(x y z) L_2(y z) L_3(z) (R^*_{12}(x))^{-1}
(R^*_{13}(x y))^{-1} (R_{23}(y))^{-1}\nonumber\\
&=& (R_{23}(y))^{-1} (R^*_{13}(x y))^{-1} (R^*_{12}(x))^{-1}
 L_3(z)  L_2(y z) L^*_1(x y z),   \label{equation}
\na
which, thought appears rather complicated, in fact is a direct consequence
of the the Yang - Baxter equation obeyed by $R$ and $R^*$ and the defining
relation (\ref{redef}) for the quantum super Yangian $\Yq$.
Equation (\ref{equation}) has various simple poles. One of them is located
at $x=q^{2(M-N)}$, the residue of which yields the following equation
\ba
& &L^*_1(y z q^{2(M-N)})  L_2(y z) L_3(z) {\bar P}_{12}[0]P_{12}
(R^*_{13}(y q^{2(M-N)}))^{-1} (R_{23}(y))^{-1}{\bar P}_{12} \nonumber \\
&=&  (R_{23}(y))^{-1}(R^*_{13}(y q^{2(M-N)}))^{-1}{\bar P}_{12}[0]P_{12}
 L_3(z)  L_2(y z) L^*_1(y z q^{2(M-N)}) {\bar P}_{12}.\label{master}
\na
To simplify this equation, we consider the factor
\ban
K_{123}(y,  z)&:=&
{\bar P}_{12}[0]P_{12}(R^*_{13}(y q^{2(M-N)}))^{-1} (R_{23}(y))^{-1}\\
 &=& (R_{23}(y))^{-1}(R^*_{13}(y q^{2(M-N)}))^{-1}{\bar P}_{12}[0]P_{12},
\nan
which appears on both sides of (\ref{master}).  An important property of
$K_{123}(y, z)$ is that
\ba
K_{123}(y,  z)(\pi^*\x\pi\x\pi)\Delta^{(2)}(u)
&=&(\pi^*\x\pi\x\pi)\Delta'^{(2)}(u)K_{123}(y,  z),\nonumber\\
& & \ \ \ \ \ \ \ \ \ \ \ \ \ \forall u\in\uq,
\label{trivial}
\na
where $\Delta^{(2)}=(\Delta\x id)\Delta$, and
$\Delta'^{(2)}=(\Delta'\x id)\Delta'$.
Since
\ban
{\bar P}[0] P (\pi^*\x\pi)\Delta(u)&=&(\pi^*\x\pi)\Delta'(u){\bar P}[0] P\\
&=&\epsilon(u), \ \ \ \ \ \forall u\in\uq,
\nan
we readily see that
\ban
[K_{123}(y,  z), \  1\x 1\x \pi(u)]&=&0,\ \ \ \ \forall u\in\uq,
\nan
and Schur's Lemma forces
\ban
K_{123}(y,  z)&=&K_{12}(y, z)\x 1,
\nan
where $K_{12}(y, z)\in End(V^*\x V)$. Using equation (\ref{trivial}) again,
we can deduce that
\ban
K_{12}(y, z)&=&f(y, z) {\bar P}_{12}[0]P_{12},
\nan
where $f(y, z)\ne 0$ is a complex function of $y$ and $z$. Inserting
$K_{123}(y,  z)$ back into (\ref{master}) leads to
\ba
L_1^*(y q^{2(M-N)})  L_2(y) {\bar P}_{12}[0]L_3(z)
&=&L_3(z){\bar P}_{12}[0] P_{12} L_2(y) L_1^*(y q^{2(M-N)}){\bar P}_{12},
\na
which, when sandwiched between two $P_{12}[0]$'s, gives rise to
\ban
[T^a_b(z), C(y)]&=&0.
\nan

Now a few remarks are in order. Set $x=q^\theta$. In the $q\rightarrow 1$
limit, $R(x)/(q-q^{-1})$ reduces to the rational solution
$S(\theta)=1 - P/\theta$ of the Yang - Baxter equation. Using $S(\theta)$,
Nazarov\cite{sy} defined the classical super Yangian associated with $\gl$,
and studied its central algebra. Our results presented here reduce to his
in the classical limit.

In studying the central algebra of the ordinary Yangian associated with
the general linear algebra $gl(M)$, a prominent role is played by the
quantum determinant, which is obtained by applying to the $M$ - fold
tensor product of the Yangian generating matrix(with appropriate choices
for the spectral parameters) the projection operator mapping the $M$ - th
rank tensor product of the $gl(M)$ vector module with itself to
the totally antisymmetric component, i.e., the trivial module\cite{molev}.
However,  no generalization of this construction to  superalgebras
seems to be possible, primarily due to the reason that no tensor product
of the $\gl$ vector module with itself contains a trivial module as an
irreducible component.  Although in our construction
of the $\Yq$ central elements we  made essential use of a trivial
$\uq$ module too, but it was manufactured by taking the tensor
product of the vector module with its dual.

Since the ordinary general linear Lie algebra $gl(M)$ is recovered
from $\gl$ by setting $N$ to zero, our treatment of $\Yq$ covers  the
ordinary quantum Yangian\cite{qy} as a special case.
It is clear that the algebra defined by equation (\ref{definition})
at $N=0$ is identical to the quantum Yangian of \cite{qy} , but the
set of invariants $C_n$ are very different from those obtained
from quantum determinants.

\section{\small UNIVERSAL L - OPERATOR AND CASIMIRS OF $\uq$}
We now examine  the super Yangian structure of  $\gl$,
then apply the results obtained to study the center of this
quantum supergroup. Let us define
\ba
l^{(+)} &=& q^{\sum_a e^a_a\x E^a_a \otimes (-1)^{[a]}}
\{ 1 \otimes 1 + (q - q^{-1}) \sum_{a<b}
e^b_a\x E^a_b (-1)^{[a]}\},\nonumber \\
l^{(-)} &=& \{ 1 \otimes 1 - (q - q^{-1}) \sum_{a<b} e^a_b\x
E^b_a (-1)^{[b]}\} q^{-\sum_a e^a_a\x  E^a_a (-1)^{[a]}},\nonumber\\
l(x)&=&\left\{l^{(+)}\ -\  x^{-1} l^{(-)}\right\}/(1-x^{-1}).
\na
Let $T: End(V)\x\uq\rightarrow \uq\x End(V)$ be the twisting map defined
by $T(e\x u)=(-1)^{[u][e]} u\x e$. Then $(1-x^{-1})T(l(x))$ coincides with
the universal $L$ operator of \cite{I}. From the results of that paper,
we can deduce that $l(x)$ satisfies the following relations
\ba
l(x)(\pi\x id)\Delta(u)&=& (\pi\x id)\Delta'(u) l(x), \ \ \ \forall u\in\uq\\
R_{12}(x) l_1(x y) l_2(y)&=& l_2(y) l_1(x y) R_{12}(x).\label{map}
\na
Note that equation (\ref{map}) is nothing else but the defining relation
of the quantum super Yangian. Thus the universal $L$ operator yields
a ${\bf Z}_2$ graded algebra homomorphism $\Lambda: \Yq\rightarrow \uq$,
\ban
\Lambda(L(x))&=& l(x).
\nan
This map is surjective, as the $E^a_b$'s generate $\uq$(in fact they form
a quantum analogue of the Cartan -- Weyl basis). Therefore, the images of
the $\Yq$ central elments $C_n$  lie in the center of $\uq$.  Set
\ban
x&=&q^\theta, \\
\Gamma&=& { {(l^{(-)})^{-1}l^{(+)}-1}\over{q - q^{-1}} },
\nan
we have
\ban
\Lambda(C(q^\theta))&=&str_{\pi}\left[\pi(q^{h_{2\rho}})
\left(\Gamma - { {q^{-\theta-2(M-N)}-1}\over{q - q^{-1}} }\right)^{-1}
\left(\Gamma - { {q^{-\theta}-1}\over{q - q^{-1}} }\right)\right].
\nan
{}From this equation we clearly see that  $\Lambda(C_n)$'s can be expressed as
 linear combinations of the following $\uq$ Casimir operators
\ban
I_n&=& str_{\pi}\left[\pi(q^{h_{2\rho}})\Gamma^n\right],
\ \ \ n=0,1,2,...,
\nan
and vice versa.  Note that the $I_n$'s are precisely the $\uq$ Casimir
operators constructed in \cite{I} using the $q$ - supertrace approach,
which bears no similarity to the Yangian construction presented here.
As pointed out in \cite{I},
this set of invariants reduces to the $\gl$ Casimirs obtained
by Jarvis and Green, and Scheunert\cite{green}, which were known to
generate the center of $\gl$.  Therefore, there is a good chance that the
$I_n$'s generate the entire center of $\uq$.

In the special case $N=0$, $\Yq$ reduces to the quantum Yangian
associated with the ordinary general linear algebra $gl(M)$,
and $I_n$'s become the $U_q(gl(M))$ Casimir operators of \cite{II}.
In this case, our arguments can be sharpened, leading to
\begin{lemma}
When  $N=0$, the $I_k$, $k=0, 1, ...$, generate the entire center of
$U_q(gl(M))$ at generic $q$.
\end{lemma}

A detailed proof of the Lemma is out of the scope of this letter,
but we can sketch the main arguments involved. Set $q=exp(t)$.
The quantum group $U_q(gl(M))$ is an associative algebra over the polynomial
ring ${\bf C}[[t]]$ completed with respect to the $t$ - adic topology;
it is also a deformation of the universal enveloping algebra $U(gl(M))$ of
$gl(M)$ in the sense of Gerstenhaber. Since the second Hochschild cohomology
group of $U(gl(M))$ with coefficients in itself is trivial, it follows that
all associative algebraic deformations of $U(gl(M))$ are trivial. This
in particular guarantees the existence of an algebra isomorphism
$F_t: U_q(gl(M))$ $\rightarrow U(gl(M))$ with
$F_t=id +t f^{(1)} + t^2 f^{(2)}+...$.  Therefore, as shown by Drinfeld,
the centers of these algebras are canonically isomorphic.
We have pointed out that $I_n(mod \  t)$ generate the center
of $U(gl(M))$, thus so do $ F_t(I_n)$, $ n=0, 1, 2, ...$.
It then follows that  $\{ I_n \ | \ n=0, 1, 2, ...\}$
is a complete set of generators for the central algebra of
$U_q(gl(M))$.  Following the same line of reasoning we can show
that the quantum Casimir operators constructed in \cite{II} generate the
centers for all quantum groups. A detailed proof of this statement will be
reported in a separate publication.

\pagebreak

\end{document}